\documentclass[english]{paper}
\usepackage[T1]{fontenc}
\usepackage[latin9]{inputenc}
\usepackage[letterpaper]{geometry}
\usepackage{float}
\usepackage{amsmath}
\usepackage{graphicx}
\usepackage{setspace}
\usepackage{amssymb}
\usepackage{esint}
\makeatletter

\providecommand{\tabularnewline}{\\}

\makeatother

\usepackage{babel}

\begin{document}

\title{Molecular zero-range potential method and its application to cyclic
structures}

\author{Dmitry V. Ponomarev%
\thanks{Currently at McMaster University, Mathematics and Statistics department.%
} , Sergey B. Leble\\{\small Gdansk University of Technology,  Theoretical Physics and Quantum Informatics department,}\\
{\small ul. Narutowicza 11/12, 80-223,  Gdansk, Poland}\\}
\maketitle
\begin{abstract}
The zero-range potentials of the radial Schrodinger equation are investigated
from a point of Darboux transformations scheme. The dressing procedure
is realized as a sequence of Darboux transformations in a way similar
to that used to obtain the generalized zero-range potentials of Huang-Derevianko
by specific choice of a family of parameters. In the present approach
we stay within the framework of conventional zero-range potential
method whilst the potential parameter (scattering length) is modified
taken into account spectral molecular properties. This allows to introduce
molecular zero-range potential once the corresponding discrete spectrum
is known. The results are illustrated on example of flat cyclic molecular
structures, with particular focus on a benzene molecule, which bounded
states energies are first found using atomic zero-range potentials,
compared with the Huckel method, and then used to introduce single
zero-range potential describing the entire molecule. Reasonable scattering
behavior for newly introduced potential gives a possibility to tackle
many-molecule problems representing molecules as appropriate single
zero-range potentials. 
\end{abstract}

\section{Introduction}

Zero-range potential (ZRP) method is proved to be an efficient tool
in describing quantum structures with arbitrary geometry at sufficiently
low energies such that detailed structure and interactions of atoms
(replaced with point-centers) between themselves can be neglected
\cite{Fermi,Breit,Albeverio,Demkov,Drukarev}. Simplicity of the approach
makes it very attractive and over the years it has been further developed
into generalized zero-range potential (gZRP) method that takes into
account effects of higher order partial waves in describing scattering
properties \cite{Huang,Huang-Yang,Derevianko,Baltenkov,Leble1,Leble2,Idziaszek}.
However, the scope of application of the regular ZRP method can be
extended to describe not only small, typically atomic, structures
but larger molecular systems. 

In the present work, after giving brief but self-consistent review
of Darboux transformation (DT) and gZRP ideas, we enhance standard
ZRP approach by means of application of a dressing procedure \cite{Doktorov,Leble1,Leble2}.
We notice and make use the fact that a certain class of DTs of spherically
symmetric ZRPs yields also a potential of zero range but, in general,
with different effective characteristic ($s$-wave scattering length
of the original ZRP is altered). Transformation parameters of a single
atomic potential can be chosen based on solution of conventional bounded
state problem solved for a system of ZRPs (molecule) such that resulting
solution mimics scattering behavior for the molecule as a whole at
low energies. 

Described ZRP treatment is demonstrated on example of a generic cyclic
structure with focus on a particular case of benzene molecule.

\section{Method of zero-range potentials}

The idea of the ZRP method is to replace real potential with a point
potential placed in the origin that mimics physical scattering behavior
in the far zone.

Therefore the equation in question is the radial Schrodinger equation
for the free space ($r>0$)\begin{equation}
-\left(\psi^{\prime\prime}+\frac{2}{r}\psi^{\prime}\right)+\frac{l\left(l+1\right)}{r^{2}}\psi=k^{2}\psi,\label{eq:Schr_eq_rad}\end{equation}
where $k^{2}=\frac{2\mu E}{\hbar^{2}}$ is wave number, $\mu$, $E$
are mass and energy of a particle (electron), respectively, $\hbar$
is the Planck's constant, $\psi=\psi(r)$ is the radial part of the
wave function.

It is well-known that the general solution can be formed as a linear
combination of spherical Bessel and Neumann functions, $j_{l}(kr)$
and $y_{l}(kr)$, respectively,\begin{equation}
\psi(r)=C_{l}\left(j_{l}(kr)-\tan\eta_{l}\cdot y_{l}(kr)\right),\label{eq:gen_sol_l}\end{equation}
or as a combination of spherical Hankel functions\begin{equation}
\psi(r)=\tilde{C}_{l}\left(s_{l}h_{l}^{(1)}(kr)-h_{l}^{(2)}(kr)\right),\label{eq:gen_sol_l_tmp}\end{equation}
with $s_{l}=\exp\left(2i\eta_{l}\right)$ being a scattering matrix.

Taking into account the following asymptotes at $kr\rightarrow0$
\cite{Abramowitz}\begin{equation}
j_{l}(kr)\approx\frac{\left(kr\right)^{l}}{\left(2l+1\right)!!},\label{eq:bess_r_small}\end{equation}

\begin{equation}
y_{l}(kr)\approx-\frac{\left(2l-1\right)!!}{\left(kr\right)^{l+1}},\label{eq:neum_r_small}\end{equation}
written with notion of the odd factorial $\left(2l+1\right)!!=\left(2l+1\right)\cdot\left(2l-1\right)\cdot\ldots\cdot3\cdot1$,
$\left(-1\right)!!=1$, we obtain asymptotic behavior of finite-energy
solution at the origin\begin{equation}
\psi(r)\approx C_{l}\left(\frac{\left(kr\right)^{l}}{\left(2l+1\right)!!}+\tan\eta_{l}\frac{\left(2l-1\right)!!}{\left(kr\right)^{l+1}}\right)\approx C_{l}\tan\eta_{l}\frac{\left(2l-1\right)!!}{\left(kr\right)^{l+1}}.\label{eq:ZRP_l_as}\end{equation}

From here, the constant $C_{l}$ can be expressed as\begin{equation}
C_{l}=\frac{\left(2l+1\right)!!}{k^{l}\left(2l+1\right)!}\left.\frac{d^{2l+1}}{dr^{2l+1}}\left(r^{l+1}\psi\right)\right|_{r=0}.\label{eq:ZRP_Cl}\end{equation}

Elimination of $C_{l}$ yields\begin{equation}
\left.\frac{1}{r^{l+1}\psi}\frac{d^{2l+1}}{dr^{2l+1}}\left(r^{l+1}\psi\right)\right|_{r=0}=\frac{\left(2l+1\right)!}{\left(2l+1\right)!!}\cdot\frac{1}{\left(2l-1\right)!!}\cdot\frac{k^{2l+1}}{\tan\eta_{l}},\label{eq:ZRP_l_cond_tmp}\end{equation}

This is to say that imposing the boundary condition 

\begin{equation}
\left.\frac{1}{r^{l+1}\psi}\frac{d^{2l+1}}{dr^{2l+1}}\left(r^{l+1}\psi\right)\right|_{r=0}=-\frac{2^{l}l!}{\left(2l-1\right)!!}1/a_{l}^{2l+1},\label{eq:ZRP_l_cond}\end{equation}
with

\begin{equation}
a_{l}^{2l+1}=-\frac{\tan\eta_{l}}{k^{2l+1}}.\label{eq:scat_length_l}\end{equation}
results in the desired physical behavior of solution at far distance
from the point-center.

Introduced quantity $a_{l}$ is termed as the partial wave scattering
length and is independent of particular energy value at low energies.
To justify this fact, it is enough to consider zero-energy solution
to the equation \eqref{eq:Schr_eq_rad}

\[
\psi(r)=A_{1}r^{l}+A_{2}\frac{1}{r^{l+1}}.\]

By matching this with \eqref{eq:gen_sol_l}, we conclude

\[
\tan\eta_{l}\approx-\underset{\text{const}}{\underbrace{A_{2}/A_{1}\dfrac{1}{\left(2l+1\right)!!\left(2l-1\right)!!}}}k^{2l+1},\]
that demonstrates energy-independence of scattering length parameters
$a_{l}$.

In most of the cases, at low energies, it is enough to consider ZRP
as a spherical point, thus neglecting higher partial waves. Therefore,
solution to the \eqref{eq:Schr_eq_rad} should be a subject to the
simple boundary condition\begin{equation}
\left.\frac{d\log\left(r\psi\right)}{dr}\right|_{r=0}=-\beta,\label{eq:ZRP_cond}\end{equation}
where $\beta=1/a_{0}$ is the $s$-wave inverse scattering length.

\section{Darboux transformation and its application}

Consider one-dimensional Sturm-Liouville equation\begin{equation}
L\psi=\lambda\psi,\label{eq:St-L_eq}\end{equation}
where \begin{equation}
L=-\dfrac{d^{2}}{dx^{2}}+u(x).\label{eq:St-L_op}\end{equation}

Now we apply such algebraic-differential transformation (referred
as Darboux transformation \cite{Darboux})

\begin{equation}
D=\dfrac{d}{dx}-\sigma(x)\label{eq:D_op}\end{equation}
that the transformed equation preserve the same form

\begin{equation}
L^{[1]}\psi^{[1]}=\lambda\psi^{[1]},\label{eq:St-L_eq_1}\end{equation}
where $\psi^{[1]}=D\psi$, $L^{[1]}=-\dfrac{d^{2}}{dx^{2}}+u^{[1]}(x)$
and the transformed potential is

\begin{equation}
u^{[1]}(x)=u(x)-2\sigma^{\prime}(x).\label{eq:u_1}\end{equation}

It is straightforward to check that \begin{equation}
\sigma(x)=\dfrac{\Phi_{1}^{\prime}}{\Phi_{1}}\label{eq:sigma_1}\end{equation}
with $\Phi_{1}$ being a particular solution to the original equation,
i.e. it satisfies \[
-\Phi_{1}^{\prime\prime}+u(x)\Phi_{1}=\lambda_{1}\Phi_{1}\]
for some $\lambda_{1}$.

It can be shown that in case of multiple application of the Darboux
transformation, the expression \eqref{eq:sigma_1} is generalized
into

\[
\sigma(x)=\frac{d}{dx}\left[\log W(\Phi_{1},\,\ldots,\,\Phi_{N})\right]\]
resulting in the following transformation of the potential

\begin{equation}
u^{[N]}(x)=u(x)-2\dfrac{d^{2}}{dx^{2}}\left[\log W(\Phi_{1},\,\ldots,\,\Phi_{N})\right].\label{eq:u_N}\end{equation}

Corresponding transformation of the solution is given by the Crum's
formula \cite{Crum,Matveev}\begin{equation}
\psi^{[N]}=D^{[N]}\psi=\dfrac{W(\Phi_{1},\,\ldots,\,\Phi_{N},\,\psi)}{W(\Phi_{1},\,\ldots,\,\Phi_{N})},\label{eq:psi_N}\end{equation}
where $W$ denote determinants of the Wronskian matrices.

Having considered one-dimensional Cartesian case, we move towards
the radial Schrodinger equation \eqref{eq:Schr_eq_rad} and notice
that it can be brought to the form \eqref{eq:St-L_eq} eligible for
direct application of obtained transformation formulas. Namely, performing
substitution $\psi=\chi/r$, one readily obtains\begin{equation}
-\chi^{\prime\prime}+\frac{l\left(l+1\right)}{r^{2}}\chi=k^{2}\chi.\label{eq:ZRP_chi_eq}\end{equation}

That is to say, that we can apply Darboux transformation to the equation
\eqref{eq:Schr_eq_rad} meaning that all original wave functions $\psi$
should be multiplied by $r$ whereas the potential term \begin{equation}
u(r)=\dfrac{l\left(l+1\right)}{r^{2}}\label{eq:u_l}\end{equation}
 remains unchanged.

Thus, the formulas \eqref{eq:u_N}, \eqref{eq:psi_N_rad} in spherical
coordinates should be replaced with\begin{equation}
u^{[N]}(r)=u(r)-2\dfrac{d^{2}}{dr^{2}}\left[\log W(r\Phi_{1},\,\ldots,\, r\Phi_{N})\right]\label{eq:u_N_rad}\end{equation}
and \begin{equation}
\psi^{[N]}=D^{[N]}\psi=\dfrac{W(r\Phi_{1},\,\ldots,\, r\Phi_{N},\, r\psi)}{rW(r\Phi_{1},\,\ldots,\, r\Phi_{N})}.\label{eq:psi_N_rad}\end{equation}

Following an idea of the previous works \cite{Doktorov}, we first
show how gZRP can be induced by application of DT.

We start by choosing a spherical Bessel function as the seed solution\begin{equation}
\psi_{l}(r)=j_{l}\left(kr\right)\label{eq:psi_l}\end{equation}
and apply $N$-th order Darboux transformation by taking spherical
Hankel functions with specific parameters $\kappa_{m}$ as prop functions

\begin{equation}
\Phi_{m}(r)=h_{l}^{(1)}(-i\kappa_{m}r),\qquad m=1,\,\ldots,\, N.\label{eq:phi_m}\end{equation}

We can employ Crum's formula \eqref{eq:psi_N} and consider asymptotic
behavior of spherical functions at $r\rightarrow\infty$\begin{equation}
j_{l}(kr)\approx\frac{\sin\left(kr-l\pi/2\right)}{kr},\label{eq:bess_r_large}\end{equation}

\begin{equation}
y_{l}(kr)\approx-\frac{\cos\left(kr-l\pi/2\right)}{kr},\label{eq:neum_r_large}\end{equation}

\begin{equation}
h_{l}^{(1)}(kr)=j_{l}(kr)+iy_{l}(kr)\approx\left(-i\right)^{l+1}\frac{\exp\left(ikr\right)}{kr},\label{eq:hank1_r_large}\end{equation}
\begin{equation}
h_{l}^{(2)}(kr)=j_{l}(kr)-iy_{l}(kr)\approx i^{l+1}\frac{\exp\left(-ikr\right)}{kr}.\label{eq:hank2_r_large}\end{equation}

Then the Wronskians turn into Vandermond determinants, hence,\begin{equation}
\psi_{l}^{[N]}(r)=C\left[\left(-i\right)^{l}\frac{\exp\left(ikr\right)}{kr}\frac{\Delta\left(\kappa_{1},\,\ldots,\,\kappa_{N},\, ik\right)}{\Delta\left(\kappa_{1},\,\ldots,\,\kappa_{N}\right)}-i^{l}\frac{\exp\left(-ikr\right)}{kr}\frac{\Delta\left(\kappa_{1},\,\ldots,\,\kappa_{N},\,-ik\right)}{\Delta\left(\kappa_{1},\,\ldots,\,\kappa_{N}\right)}\right],\label{eq:psi_l_N_as}\end{equation}
where $C$ represents here and later on a generic constant without
prescribing it any specific value such that it can absorb constant
multipliers where their meaning is not important without changing
notation.

We note that for $l=0$ this expression is not asymptotic, but exact.

A Vandermond determinant in \eqref{eq:psi_l_N_as} can be computed
by noticing that $k=-i\kappa_{m}$ (for $m=1,\,\ldots,\, N$) are
the roots of polynomial with respect to $k$ equation that is obvious
due to the fact that replacement $ik\rightarrow\kappa_{m}$ yields
zero determinant due to linear dependencies of the rows, thereby allowing
the following factorization \[
\Delta\left(\kappa_{1},\,\ldots,\,\kappa_{N},\, ik\right)=\left|\begin{array}{ccccc}
1 & \kappa_{1} & \kappa_{1}^{2} & \dots & \kappa_{1}^{N}\\
1 & \kappa_{2} & \kappa_{2} & \dots & \kappa_{2}^{N}\\
\dots & \dots & \dots & \dots & \dots\\
1 & \kappa_{N} & \kappa_{N}^{2} & \dots & \kappa_{N}^{N}\\
1 & ik & \left(ik\right)^{2} & \dots & \left(ik\right)^{N}\end{array}\right|=C\prod_{m=1}^{N}\left(\kappa_{m}-ik\right).\]

Denoting

\begin{equation}
s_{l}=\prod_{m=1}^{N}\frac{\left(\kappa_{m}-ik\right)}{\left(\kappa_{m}+ik\right)}=-\prod_{m=1}^{N}\frac{\left(k+i\kappa_{m}\right)}{\left(k-i\kappa_{m}\right)},\label{eq:Darboux_s_l}\end{equation}
we recognize in the expression \eqref{eq:psi_l_N_as} the asymptotes
of spherical Hankel functions, hence\begin{equation}
\psi_{l}^{[N]}(r)=C\left[s_{l}h_{l}^{(1)}(kr)-h_{l}^{(2)}(kr)\right].\label{eq:psi_l_N_final}\end{equation}

As one can notice, the expression \eqref{eq:psi_l_N_final} coincides
with \eqref{eq:gen_sol_l_tmp} if \begin{equation}
\exp\left(2i\eta_{l}\right)=\prod_{m=1}^{N}\frac{\left(\kappa_{m}-ik\right)}{\left(\kappa_{m}+ik\right)},\label{eq:Darboux_ZRP_cond}\end{equation}
or, taking into account \eqref{eq:scat_length_l},\begin{equation}
\tan\eta_{l}=-a_{l}^{2l+1}k^{2l+1}=-i\frac{\prod_{m=1}^{N}\left(\kappa_{m}-ik\right)-\prod_{m=1}^{N}\left(\kappa_{m}+ik\right)}{\prod_{m=1}^{N}\left(\kappa_{m}-ik\right)+\prod_{m=1}^{N}\left(\kappa_{m}+ik\right)}.\label{eq:Darboux_ZRP_cond_tmp1}\end{equation}

Thus we conclude that for the direct correspondence to the gZRP behavior,
the number of consequent transformations $N=2l+1$ should be taken.

Given some quantity $a=\left|a\right|e^{i\phi_{a}}$, the parameters
$\kappa_{1},\,\ldots,\,\kappa_{2l+1}$ needs to be chosen in a way
that \begin{equation}
\prod_{m=1}^{2l+1}\left(\kappa_{m}+ik\right)=\left(ik\right)^{2l+1}-a.\label{eq:prod_plus_ik}\end{equation}

This is equivalent to the $ik=-\kappa_{m}$ ($m=1,\,\ldots,\,2l+1$)
being the roots of the equation \[
\left(ik\right)^{2l+1}=a,\]
that is to say $-\kappa_{m}=\sqrt[\left(2l+1\right)]{a}$, or \begin{equation}
-\kappa_{m}=\left|a\right|^{1/\left(2l+1\right)}\exp\left(i\frac{\left(\phi_{a}+2\pi m\right)}{2l+1}\right),\qquad m=1,\,\ldots,\,2l+1.\label{eq:kappa_m}\end{equation}

In the similar manner we obtain\begin{equation}
\prod_{m=1}^{2l+1}\left(\kappa_{m}-ik\right)=\left(-ik\right)^{2l+1}-a.\label{eq:prod_minus_ik}\end{equation}

Substitution of \eqref{eq:prod_plus_ik}, \eqref{eq:prod_minus_ik}
into \eqref{eq:Darboux_ZRP_cond_tmp1} results in\[
a=\left(-1\right)^{l+1}/a_{l}^{2l+1}.\]

Therefore, providing $a_{l}$ is a real number, \eqref{eq:kappa_m}
yields\begin{equation}
\kappa_{m}=-1/a_{l}\cdot\exp\left(i\pi\frac{l+2m+1}{2l+1}\right),\qquad m=1,\,\ldots,\,2l+1.\label{eq:kappa_m_final}\end{equation}

Now alternatively to gZRP instead of taking into account impact of
higher harmonics, we consider the transformation of spherically symmetric
solution (i.e. $l=0$) with the parameters $\kappa_{m}$ to be chosen
according to spectral data of the entire molecule obtained, for example,
from the discrete spectrum of molecule modeled by regular ZRP method.

We get back to \eqref{eq:Darboux_ZRP_cond_tmp1} and since \[
\prod_{m=1}^{N}\left(\kappa_{m}+ik\right)=\prod_{m=1}^{N}\kappa_{m}+ik\sum_{n=1}^{N}\prod_{\underset{m\neq n}{m=1}}^{N}\kappa_{m}+\left(ik\right)^{2}\sum_{j=1}^{N}\sum_{n<j}^{N}\prod_{\underset{m\neq n,\, m\neq j}{m=1}}^{N}\kappa_{m}+\ldots+\left(ik\right)^{N-1}\sum_{n=1}^{N}\kappa_{n}+\left(ik\right)^{N},\]
we continue the last equality as\begin{equation}
a_{l}^{2l+1}k^{2l+1}=-i\frac{ik\sum_{n=1}^{N}\prod_{\underset{m\neq n}{m=1}}^{N}\kappa_{m}+\ldots+\left(ik\right)^{N}}{\prod_{m=1}^{N}\kappa_{m}+\ldots+\left(ik\right)^{N-1}\sum_{n=1}^{N}\kappa_{n}}.\label{eq:Darboux_ZRP_cond_tmp2}\end{equation}

Limiting ourselves with $l=0$ case, we consider low-energy limit
neglecting terms of higher order than one with respect to $k$ to
obtain\begin{equation}
a_{0}=\frac{\sum_{n=1}^{N}\prod_{\underset{m\neq n}{m=1}}^{N}\kappa_{m}}{\prod_{m=1}^{N}\kappa_{m}}.\label{eq:a0}\end{equation}

Alternatively, recalling notation $\beta=1/a_{0}$, the last expression
can be rewritten as\begin{equation}
\beta=\left(\sum_{m=1}^{N}\kappa_{m}^{-1}\right)^{-1}\label{eq:beta_eff}\end{equation}

On the other hand, we can readily observe that Darboux transformation
of the seed solution \eqref{eq:psi_l} with the prop functions \eqref{eq:phi_m}
results in the solution corresponding to the ZRP for arbitrary choice
of $N$ and parameters $\kappa_{m}$, $m=1,\,\ldots,\, N$. Indeed,
for $l=0$ \[
\log W(r\Phi_{1},\,\ldots,\, r\Phi_{N})=\sum_{m=1}^{N}\kappa_{m}r+C,\]
thus, according to \eqref{eq:u_N_rad},\[
u^{[N]}(r)\equiv0,\qquad r>0.\]

Freedom of choice of transformation parameters gives a possibility
to induce desired poles of scattering matrix \eqref{eq:Darboux_s_l}
and, thereby, perform transition from atomic to molecular ZRP by choosing
$\kappa_{m}$, $m=1,\,\ldots,\, M$ such that new potential allows
$M$ bounded states obtained by solving conventional formulation of
the discrete spectrum problem for the set of $M$ standard ZRPs making
up the molecule. Molecular ZRP can be qualitatively characterized
by effective scattering length computed according to \eqref{eq:a0}
and thus used in such simplified scattering calculations in complicated
problems involving, for example, a chain of molecules.

Below we demonstrate the presented approach by applying it to a general
cyclic molecule with eventual focus on a benzene molecule.

\section{Case study: a cyclic molecule}

In order to describe a cyclic structure, we consider a system (molecule)
of $M$ point-centers, typically atoms, modeled by ZRPs which positions
are kept fixed in the plane in space in a cyclic order.

\begin{figure}[H]

\centering{}\includegraphics[scale=0.6]{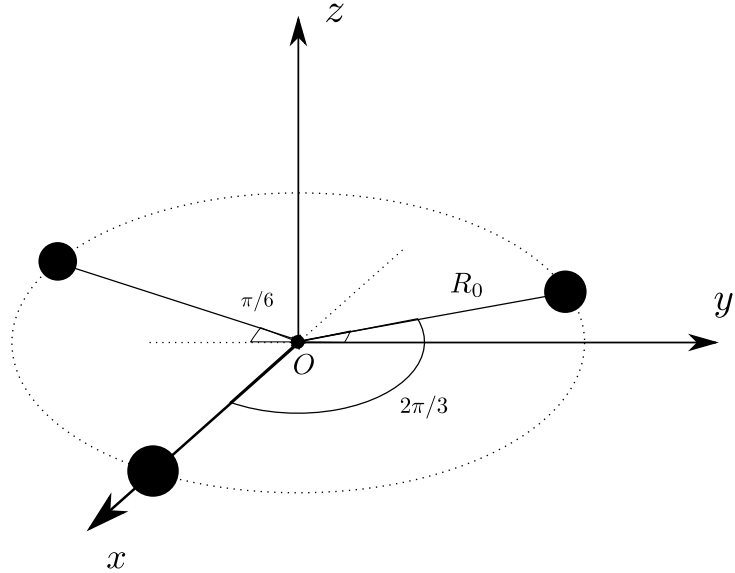}\caption{\label{fig:geom_3at} Illustration of the geometry by an example of
3 atoms.}

\end{figure}

Spatial coordinates are chosen such that all atoms lie in the $xy$-plane
with the origin coinciding with the center of the molecule. Atoms
are numerated counterclockwise and the $x$-axis is chosen along the
direction to the first atom. Described geometry is illustrated on
the Fig. \ref{fig:geom_3at}.

\vspace{3mm}

\vspace{3mm}

\subsection{Bounded state problem}

According to the idea of ZRP method, we write solution to the bounded
state problem as

\begin{equation}
\psi(\vec{r})=\sum_{j=1}^{M}C_{j}\cdot g\left(\left|\vec{r}-\vec{R}_{j}\right|,\,\kappa\right),\label{eq:ZRP_CM_sol}\end{equation}
subject to the conditions \eqref{eq:ZRP_cond} 

\begin{equation}
\left.\dfrac{\partial\log\left(\left|\vec{r}-\vec{R}_{i}\right|\cdot\psi(\vec{r})\right)}{\partial\left|\vec{r}-\vec{R}_{i}\right|}\right|_{\left|\vec{r}-\vec{R}_{i}\right|=0}=-\beta,\qquad i=1,\,\dots,\, M,\label{eq:ZRP_CM_cond}\end{equation}
where \begin{equation}
g\left(r,\,\kappa\right)=\dfrac{e^{-\kappa r}}{r}.\label{eq:single_bound_l0}\end{equation}

\begin{figure}[H]

\begin{centering}
\includegraphics[scale=0.5]{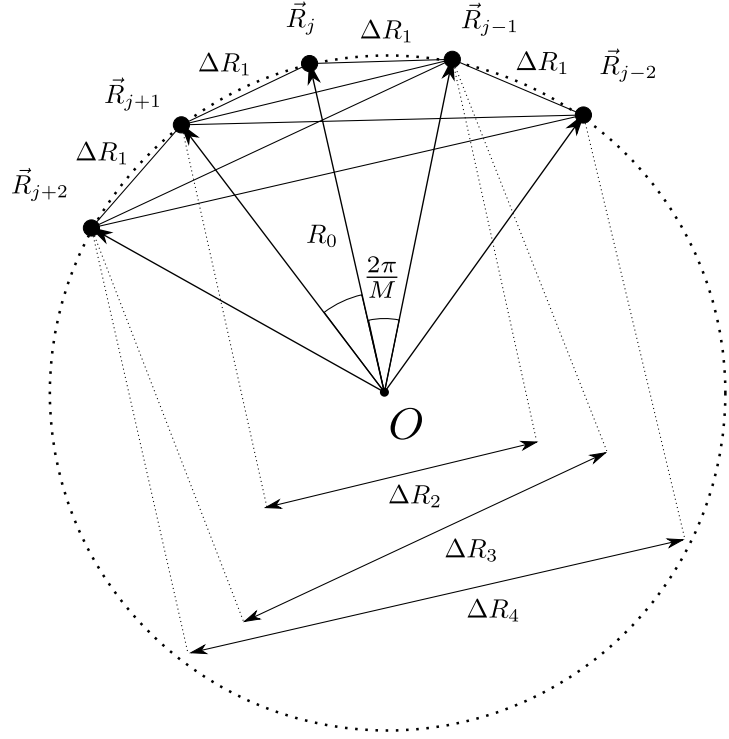}
\par\end{centering}

\caption{\label{fig:geom_dist} Illustration of the distance notation.}

\end{figure}

Let us introduce the following notation. We will denote a distance
between nearest neighboring atoms as $\Delta R_{1}$, between every
second atom from each given one as $\Delta R_{2}$, every third as
$\Delta R_{3}$, and so on (see the Fig. \ref{fig:geom_dist}).

\vspace{3mm}

\vspace{3mm}

From simple geometrical considerations it follows that \begin{equation}
\Delta R_{j}=2R_{0}\sin\left(\dfrac{2\pi}{M}\cdot\dfrac{j}{2}\right)=2R_{0}\sin\left(\phi_{j+1}/2\right),\label{eq:delta_R_j}\end{equation}
where $\phi_{j}=\dfrac{2\pi}{M}\left(j-1\right)$ is azimuthal angular
coordinate of a $j$-th atom, $R_{0}$ is radius of the molecule.
We note that the expression is formally valid for $j=1,\,\dots,\, M$,
however we note that there is only $\left[M/2\right]$ different values
since $\Delta R_{\left[M/2\right]-j}=\Delta R_{\left[M/2\right]+j}$
for $j=1,\,\dots,\,\left[M/2\right]-1$ where $[\cdot]$, here and
later on, marks integer part of an argument (i.e. the floor function).

We develop conditions \eqref{eq:ZRP_CM_cond} to show existence of
the derivatives\begin{eqnarray*}
\left.\dfrac{\partial\log\left(\left|\vec{r}-\vec{R}_{i}\right|\cdot\psi(\vec{r})\right)}{\partial\left|\vec{r}-\vec{R}_{i}\right|}\right|_{\left|\vec{r}-\vec{R}_{i}\right|=0} & = & \dfrac{1}{C_{i}}\left[\left(\ldots+C_{i-1}\cdot g\left(\Delta R_{1},\,\kappa\right)-C_{i}\cdot\kappa+C_{i+1}\cdot g\left(\Delta R_{1},\,\kappa\right)+\ldots\right)+\right.\\
 &  & \left.+\left.\left|\vec{r}-\vec{R}_{i}\right|\cdot\dfrac{\partial\left(\psi(\vec{r})-g\left(\left|\vec{r}-\vec{R}_{i}\right|,\,\kappa\right)\right)}{\partial\left|\vec{r}-\vec{R}_{i}\right|}\right|_{\left|\vec{r}-\vec{R}_{i}\right|=0}\right].\end{eqnarray*}

To show that the last term in the square bracketed expression is well-defined
and equal to zero due to the presence of the multiplier $\left|\vec{r}-\vec{R}_{i}\right|$,
we first note that $\psi(\vec{r})-g\left(\left|\vec{r}-\vec{R}_{i}\right|,\,\kappa\right)$
is a function of the arguments $\left|\vec{r}-\vec{R}_{j}\right|$
for $j=1,\,\dots,\, N$, $j\neq i$, which does not lead to a singularity
when computed at $\left|\vec{r}-\vec{R}_{i}\right|=0$. Thus, due
to the chain rule, it remains to demonstrate that the derivatives
$\frac{\partial\left|\vec{r}-\vec{R}_{j}\right|}{\partial\left|\vec{r}-\vec{R}_{i}\right|}$
do not blow up at $\left|\vec{r}-\vec{R}_{i}\right|=0$. This can
be done in a straightforward manner as it follows

\begin{eqnarray*}
\frac{\partial\left|\vec{r}-\vec{R}_{j}\right|}{\partial\left|\vec{r}-\vec{R}_{i}\right|} & = & \frac{\partial\left|\vec{r}-\vec{R}_{i}+\Delta\vec{R}_{ij}\right|}{\partial\left|\vec{r}-\vec{R}_{i}\right|}=\frac{\partial\,\sqrt{\left|\vec{r}-\vec{R}_{i}\right|^{2}+2\cdot\left(\Delta\vec{R}_{ij},\,\vec{r}-\vec{R}_{i}\right)+\left|\Delta\vec{R}_{ij}\right|^{2}}}{\partial\left|\vec{r}-\vec{R}_{i}\right|}=\\
 & = & \frac{1}{\left|\vec{r}-\vec{R}_{j}\right|}\left(\left|\vec{r}-\vec{R}_{i}\right|+\frac{d\left(\Delta\vec{R}_{ij},\,\vec{r}-\vec{R}_{i}\right)}{d\left|\vec{r}-\vec{R}_{i}\right|}\right),\end{eqnarray*}
where $\Delta\vec{R}_{ij}=\vec{R}_{i}-\vec{R}_{j}$.

Taking into account that

\[
\left.\frac{d\left(\Delta\vec{R}_{ij},\,\vec{r}-\vec{R}_{i}\right)}{d\left|\vec{r}-\vec{R}_{i}\right|}\right|_{\left|\vec{r}-\vec{R}_{i}\right|=0}=\lim_{\left|\vec{\delta}\right|\rightarrow0}\frac{\left(\Delta\vec{R}_{ij},\,\vec{\delta}\right)-0}{\left|\vec{\delta}\right|}=\left|\Delta\vec{R}_{ij}\right|\cdot\cos\alpha_{ij},\]
where $\cos\alpha_{ij}=\frac{\left(\Delta\vec{R}_{ij},\,\vec{\delta}\right)}{\left|\Delta\vec{R}_{ij}\right|\cdot\left|\vec{\delta}\right|}$,
we conclude

\[
\left.\frac{\partial\left|\vec{r}-\vec{R}_{j}\right|}{\partial\left|\vec{r}-\vec{R}_{i}\right|}\right|_{\left|\vec{r}-\vec{R}_{i}\right|=0}=\cos\alpha_{ij},\]

thus, we have shown the required boundedness\[
-1\leq\left.\frac{\partial\left|\vec{r}-\vec{R}_{j}\right|}{\partial\left|\vec{r}-\vec{R}_{i}\right|}\right|_{\left|\vec{r}-\vec{R}_{i}\right|=0}\leq1.\]

Therefore, eventually,\[
\left.\dfrac{\partial\log\left(\left|\vec{r}-\vec{R}_{i}\right|\cdot\psi(\vec{r})\right)}{\partial\left|\vec{r}-\vec{R}_{i}\right|}\right|_{\left|\vec{r}-\vec{R}_{i}\right|=0}=\dfrac{1}{C_{i}}\left(\ldots+C_{i-1}\cdot g\left(\Delta R_{1},\,\kappa\right)-C_{i}\cdot\kappa+C_{i+1}\cdot g\left(\Delta R_{1},\,\kappa\right)+\ldots\right)=-\beta\]
for $i=1,\,\ldots,\, M$.

This gives the set of homogeneous linear algebraic equations allowing
to determine $C_{1},\,\ldots,\, C_{M}$ while its compatibility condition
yields spectral equation with respect to $\kappa$.

Alternatively, the set of constants can be found from general symmetrical
considerations and spectral condition will follow from their direct
substitution into the last equation taken, for instance, at $i=1$.
Thus the spectral condition reads

\begin{equation}
\beta-\kappa+\sum_{j=1}^{M-1}C_{j+1}/C_{1}\cdot g\left(\Delta R_{j},\,\kappa\right)=0.\label{eq:spectr_cond_tmp}\end{equation}

To find all constants $C_{j}$ employing symmetry of the problem,
we introduce cyclic permutation operator $T$ which action on the
wave function \eqref{eq:ZRP_CM_sol} results in the replacement $\vec{R}_{1}\rightarrow\vec{R}_{2},\,\vec{R}_{2}\rightarrow\vec{R}_{3},\,\ldots,\,\vec{R}_{M}\rightarrow\vec{R}_{1}$.
Since $T^{M}\psi(\vec{r})=\psi(\vec{r})$, we immediately find the
operator eigenvalues \begin{equation}
\lambda_{n}=\exp\left(2\pi in/M\right),\qquad n=0,\,\ldots,\, M-1,\label{eq:lambda_n}\end{equation}
where $i$ here, and onwards in the similar combinations, stands for
the imaginary unit $i=\sqrt{-1}$.

To proceed with corresponding set of constants, as a calculation trick,
we assume an observation point to be arbitrary but close to the center
of the molecule i.e. $r\ll R_{0}$ such that we can do approximation
$e^{-\kappa\left|\vec{r}-\vec{R}_{j}\right|}\approx e^{-\kappa R_{0}}$
and write the following expansions

\begin{equation}
\dfrac{1}{\left|\vec{r}-\vec{R}_{j}\right|}=\sum_{l=0}^{\infty}\dfrac{r^{l}}{R_{0}^{l+1}}P_{l}\left(\cos\gamma_{j}\right),\qquad j=1,\,\ldots,\, M,\label{eq:Leg_expan}\end{equation}
where $\cos\gamma_{j}=\dfrac{\vec{r}\cdot\vec{R_{j}}}{\left|\vec{r}\right|\cdot R_{0}}$.

The addition theorem for spherical harmonics (e.g. \cite{Arfken})
allows to write\begin{equation}
P_{l}\left(\cos\gamma_{j}\right)=\dfrac{4\pi}{2l+1}\sum_{m=-l}^{l}(-1)^{m}\cdot Y_{l}^{m}(\theta,\,\phi)\cdot Y_{l}^{-m}(\pi/2,\,0)\cdot e^{-i\left(j-1\right)\frac{2\pi m}{M}}\label{eq:Leg_addit}\end{equation}

Substitution of \eqref{eq:ZRP_CM_sol}, \eqref{eq:lambda_n}-\eqref{eq:Leg_addit}
into the eigenvalue problem\[
T\psi(\vec{r})=\lambda_{n}\psi(\vec{r})\]
yields\[
\sum_{m}\sum_{j=1}^{M}C_{j}e^{-ik\frac{2\pi m}{M}}=e^{i\frac{2\pi n}{M}}\sum_{m}\sum_{j=1}^{M}C_{j}e^{-i\left(j-1\right)\frac{2\pi m}{M}},\]
that is equivalent to\[
\begin{cases}
\sum_{m}\sum_{j=1}^{M}C_{j}e^{-ij\frac{2\pi m}{M}}=e^{i\frac{2\pi n}{M}}\sum_{m}\sum_{j=1}^{M}C_{j+1}e^{-ij\frac{2\pi m}{M}}\\
C_{M+1}=C_{1}\end{cases}.\]

This allows to conclude\[
C_{j+1}=e^{-i\frac{2\pi n}{M}}C_{j},\]
and, consequently,\begin{equation}
C_{j+1}=e^{-ij\frac{2\pi n}{M}}C_{1},\qquad j=1,\,\ldots,\, M-1,\qquad n=0,\,\ldots,\, M-1.\label{eq:Cj_sol}\end{equation}

Feeding this into \eqref{eq:spectr_cond_tmp} solves the problem of
finding energy levels \begin{equation}
E_{n}=-\frac{\hbar^{2}\kappa_{n}^{2}}{2\mu},\qquad n=0,\,\ldots,\,\left[M/2\right].\label{eq:E_n}\end{equation}
 by treating transcendental equation with respect to $\kappa_{n}$
(by means of numerical methods, e.g. the Newton-Raphson algorithm)

\begin{equation}
\beta-\kappa_{n}+\sum_{j=1}^{M}e^{-ij\frac{2\pi n}{M}}\cdot g\left(\Delta R_{j},\,\kappa_{n}\right)=0.\label{eq:spect_cond}\end{equation}

Structure of the spectrum is as it follows. Ground state $n=0$ and
the state with highest energy $n=M/2$ (in case of even number of
atoms $M$) are the only non-degenerate ones,\begin{equation}
\psi_{0}(\vec{r})=C_{1}\sum_{j=1}^{M}g\left(\left|\vec{r}-\vec{R}_{j}\right|,\,\kappa\right),\label{eq:spect_psi_0}\end{equation}
\begin{equation}
\psi_{M/2}(\vec{r})=C_{1}\sum_{j=1}^{M}\left(-1\right)^{j-1}\cdot g\left(\left|\vec{r}-\vec{R}_{j}\right|,\,\kappa\right),\label{eq:spect_psi_half}\end{equation}
whilst all the other states are doubly-degenerate with complex conjugated
wave functions\begin{equation}
\psi_{n}(\vec{r})=C_{1}\sum_{j=1}^{M}e^{-i\left(j-1\right)\frac{2\pi n}{M}}\cdot g\left(\left|\vec{r}-\vec{R}_{j}\right|,\,\kappa\right),\label{eq:spect_psi_n}\end{equation}
\begin{equation}
\psi_{\bar{n}}(\vec{r})=C_{1}\sum_{j=1}^{M}e^{i\left(j-1\right)\frac{2\pi n}{M}}\cdot g\left(\left|\vec{r}-\vec{R}_{j}\right|,\,\kappa\right),\qquad n=1,\,\ldots,\,\left[M/2\right]-1,\label{eq:spect_psi_n_conj}\end{equation}
where the constant $C_{1}$ can be found from the normalization condition.

\subsection{Continuous spectrum problem}

Now we consider the scattering of plane wave incident axially on the
$M$-atom potential plane symmetrical structure under question.

We write the solution to the Schrodinger equation as \begin{equation}
\psi(\vec{r})=A_{0}e^{ikz}+\sum_{j=1}^{M}C_{j}\cdot g\left(\left|\vec{r}-\vec{R}_{j}\right|,\,-ik\right),\label{eq:scat_ZRP_gen_sol}\end{equation}
where $k=\sqrt{\dfrac{2\mu E}{\hbar^{2}}}$, $A_{0}$ is an amplitude
of the incident plane wave and $g\left(r,\,\kappa\right)$ is as introduced
above according to \eqref{eq:single_bound_l0}.

As before (and using previously described notation), at each atom
the solution must satisfy the ZRP condition \eqref{eq:ZRP_cond}

\[
\left.\dfrac{\partial\log\left(\left|\vec{r}-\vec{R}_{j}\right|\cdot\psi(\vec{r})\right)}{\partial\left|\vec{r}-\vec{R}_{j}\right|}\right|_{\left|\vec{r}-\vec{R}_{j}\right|=0}=\dfrac{1}{C_{j}}\left(A_{0}+\ldots+C_{j-1}\cdot g\left(\Delta R_{1},\,-ik\right)+C_{j}\cdot ik+C_{j+1}\cdot g\left(\Delta R_{1},\,-ik\right)+\ldots\right)=-\beta.\]

This results in the set of linear equations\[
\left(\begin{array}{ccccc}
\beta+ik & g\left(\Delta R_{1},\,-ik\right) & \dots & g\left(\Delta R_{2},\,-ik\right) & g\left(\Delta R_{1},\,-ik\right)\\
g\left(\Delta R_{1},\,-ik\right) & \beta+ik & \dots & g\left(\Delta R_{3},\,-ik\right) & g\left(\Delta R_{2},\,-ik\right)\\
\dots & \dots & \ddots & \dots & \dots\\
g\left(\Delta R_{2},\,-ik\right) & g\left(\Delta R_{3},\,-ik\right) & \dots & \beta+ik & g\left(\Delta R_{1},\,-ik\right)\\
g\left(\Delta R_{1},\,-ik\right) & g\left(\Delta R_{2},\,-ik\right) & \dots & g\left(\Delta R_{1},\,-ik\right) & \beta+ik\end{array}\right)\left(\begin{array}{c}
C_{1}\\
C_{2}\\
\vdots\\
C_{M-1}\\
C_{M}\end{array}\right)=-A_{0}\left(\begin{array}{c}
1\\
1\\
\vdots\\
1\\
1\end{array}\right),\]

that obviously has the solution corresponding to trivial symmetry
of the system

\begin{equation}
C_{1}=\ldots=C_{M}\equiv C_{0M}.\label{eq:scat_C_sol}\end{equation}

From any of the equation of the set it follows that\[
C_{0M}=\begin{cases}
-\dfrac{A_{0}}{\beta+ik+2\sum_{j=1}^{\left(M-1\right)/2}g\left(\Delta R_{j},\,-ik\right)}, & M\,-\,\text{odd},\\
-\dfrac{A_{0}}{\beta+ik+2\sum_{j=1}^{M/2-1}g\left(\Delta R_{j},\,-ik\right)+g\left(\Delta R_{M/2},\,-ik\right)}, & M\,-\,\text{even.}\end{cases}\]

This can be unified as \begin{equation}
C_{0M}=-\dfrac{A_{0}}{\beta+ik+2\sum_{j=1}^{\left(M+\chi_{M}\right)/2-1}g\left(\Delta R_{j},\,-ik\right)+\chi_{M-1}\cdot g\left(\Delta R_{M/2},\,-ik\right)},\label{eq:scat_C0M}\end{equation}
where $\chi_{M}=\frac{M/2-\left[M/2\right]}{M/2-\left[\left(M-1\right)/2\right]}$
is an indicator of parity of the number of atoms $M$.

It is also useful to recall \eqref{eq:delta_R_j} and rewrite the
sum explicitly\begin{equation}
\sum_{j=1}^{\left(M+\chi_{M}\right)/2-1}g\left(\Delta R_{j},\,-ik\right)=\frac{1}{2R_{0}}\sum_{j=2}^{\left(M+\chi_{M}\right)/2}\frac{\exp\left(2ikR_{0}\sin\left(\phi_{j}/2\right)\right)}{\sin\left(\phi_{j}/2\right)}\label{eq:C0M_calc_l0}\end{equation}
with $\phi_{j}=\frac{2\pi}{M}\left(j-1\right)$.

Having obtained that, and taking into account the expansion $\left|\vec{r}-\vec{R}_{j}\right|\approx r-\frac{\vec{r}}{r}\cdot\vec{R_{j}}$
at $r\rightarrow\infty$, we can present the solution \eqref{eq:scat_ZRP_gen_sol}
as

\begin{equation}
\psi(\vec{r})=A_{0}\left(e^{ikz}+f(\theta,\,\phi)\dfrac{e^{ikr}}{r}\right),\label{eq:scat_ZRP_gen_sol_as}\end{equation}
where the scattering amplitude at long distances in direction $\vec{n}=\vec{r}/r=\left(\sin\theta\cos\phi,\,\sin\theta\sin\phi,\,\cos\theta\right)^{T}$
is given by\begin{equation}
f(\theta,\,\phi)=C_{0M}/A_{0}\sum_{j=1}^{M}e^{-ik\vec{n}\cdot\vec{R_{j}}}=C_{0M}/A_{0}\sum_{j=1}^{M}\exp\left\{ -ikR_{0}\sin\theta\cos\left(\phi-\phi_{j}\right)\right\} .\label{eq:scat_ampl}\end{equation}

From here the differential cross-section can be determined\begin{equation}
\frac{d\sigma}{d\Omega}=\left|f(\theta,\,\phi)\right|^{2},\label{eq:diff_cross_sect}\end{equation}
as well as the total scattering cross-section\begin{equation}
\sigma=\int_{0}^{2\pi}\int_{0}^{\pi}\left|f(\theta,\,\phi)\right|^{2}\sin\theta d\theta d\phi.\label{eq:int_cross_sect}\end{equation}

This also can be expressed according to the optical theorem\begin{equation}
\sigma=\frac{4\pi}{k}\text{Im}\left\{ f(0,\,\phi)\right\} .\label{eq:opt_theorem}\end{equation}

Therefore, \begin{eqnarray}
\sigma & = & \frac{4\pi M}{k}\text{Im}\left\{ C_{0M}/A_{0}\right\} =\frac{4\pi M}{k}\frac{k+P}{\left(k+P\right)^{2}+\left(\beta+Q\right)^{2}},\label{eq:int_cross_sect_final}\end{eqnarray}
 where we denote

\begin{equation}
P\equiv\frac{1}{R_{0}}\cdot\left\{ \sum_{j=1}^{\left(M+\chi_{M}\right)/2-1}\frac{\sin\left(2kR_{0}\sin\left(\pi j/M\right)\right)}{\sin\left(\pi j/M\right)}+\frac{\chi_{M-1}}{2}\cdot\sin\left(2kR_{0}\right)\right\} ,\label{eq:P_l0}\end{equation}

\begin{equation}
Q\equiv\frac{1}{R_{0}}\cdot\left\{ \sum_{j=1}^{\left(M+\chi_{M}\right)/2-1}\frac{\cos\left(2kR_{0}\sin\left(\pi j/M\right)\right)}{\sin\left(\pi j/M\right)}+\frac{\chi_{M-1}}{2}\cdot\cos\left(2kR_{0}\right)\right\} .\label{eq:Q_l0}\end{equation}

We note that for the case of a single ZRP scatterer (i.e. $M=1$)
characterized by\begin{equation}
\left.\dfrac{d\log\left(r\psi(r)\right)}{dr}\right|_{r=0}=-\beta_{0},\label{eq:ZRP_C1_cond}\end{equation}
the formula \eqref{eq:scat_C0M} should be obviously replaced with\begin{equation}
C_{01}=-\frac{A_{0}}{\beta_{0}+ik}\label{eq:scat_C01}\end{equation}
resulting in the following formula for total scattering cross-section
\begin{equation}
\sigma=\frac{4\pi}{k}\text{Im}\left\{ C_{01}/A_{0}\right\} =\frac{4\pi}{k^{2}+\beta_{0}^{2}}.\label{eq:int_cross_sect_single}\end{equation}

As it was announced, we believe that a simplified theoretical model
of molecular structures can be built by replacing each multiatomic
molecule with some ZRP once its spectrum is known. Indeed, in particular
for scattering on one molecule instead of \eqref{eq:int_cross_sect_final}
one would attempt to use \eqref{eq:int_cross_sect_single} with $\beta_{0}$
to be chosen from the discrete spectrum levels $\kappa_{m}$ according
to the obtained formula \eqref{eq:beta_eff}.

\subsection{Results for benzene molecule}

Focusing on particular case of a benzene molecule $C_{6}H_{6}$ that
can be considered as a cyclic structures formed by $C-H$ constituents
described by ZRPs, we set $M=6$ and start with the bounded state
problem solution.

\begin{figure}[H]
\centering{}\includegraphics[scale=0.4, angle=270]{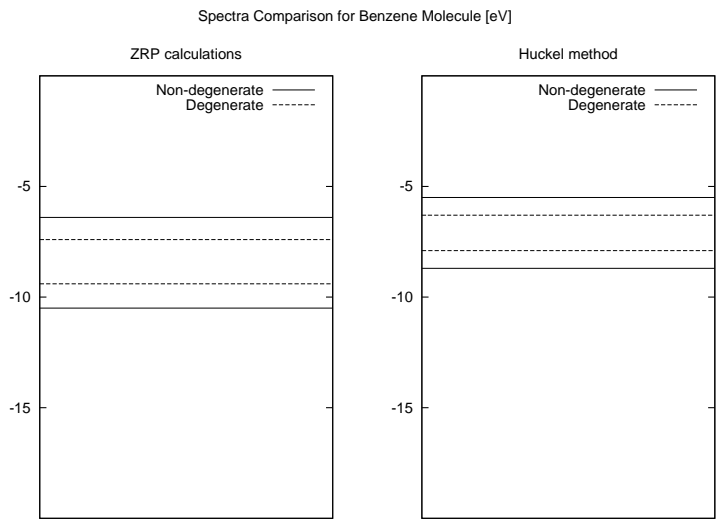}\caption{\label{fig:ZRP_Huck_comp} Comparison of ZRP and Huckel calculations
of energy levels for benzene molecule}

\end{figure}

Structure of the discrete spectrum is as described by \eqref{eq:spect_psi_0}-\eqref{eq:spect_psi_n_conj},
i.e. there are 4 energy levels with 2 middle ones being doubly degenerate.
Numerical solution of the transcendental equations \eqref{eq:spect_cond}
yields the values of energy levels which are compared with those obtained
by the well-known Huckel method on the Fig. \ref{fig:ZRP_Huck_comp}.
The spectrum, namely distances between the energy levels, reveals
good correspondence of ZRP approach to the qualitative Huckel calculations
dictating energies from the ground to the highest level to be $E_{0}-2A$,
$E_{0}-A$, $E_{0}+A$, $E_{0}+2A$, respectively \cite{LeBellac,Coulson}.

\vspace{3mm}

\vspace{3mm}

To analyze continuous spectrum, we consider total scattering cross-section
dependence on energy.

We compare the results given by the standard (atomic) ZRP model of
the molecule \eqref{eq:int_cross_sect_final} and those followed from
single ZRP description of the molecule, i.e. \eqref{eq:int_cross_sect_single}
using the inverse of effective scattering length as introduced by
\eqref{eq:beta_eff}. As one can observe on the plots given below
for differently estimated values of radius of benzene molecule (e.g.
see the introduction of \cite{Nakahara}), there is moderate coincidence
between conventional set of atomic ZRPs model and proposed single
molecular ZRP approach.

\begin{minipage}[t]{0.45\columnwidth}%
\begin{center}
\begin{figure}[H]
\includegraphics[scale=0.3, angle=270]{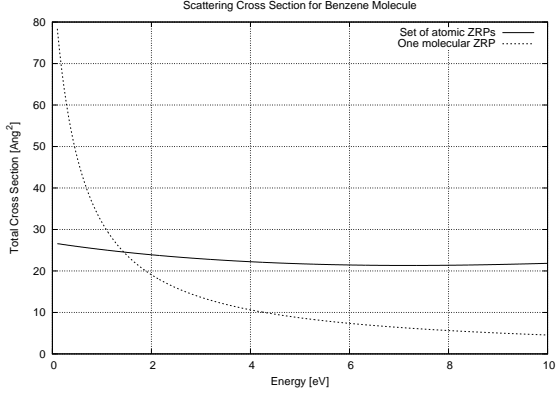}\caption{\label{fig:sigma_R01} Total scattering cross-sections for atomic
and molecular ZRPs: smaller radius $R_{01}$ of benzene molecule ($C-C$
bond length estimation).}

\end{figure}

\par\end{center}%
\end{minipage} $\qquad$%
\begin{minipage}[t]{0.45\columnwidth}%
\begin{center}
\begin{figure}[H]
\includegraphics[scale=0.3, angle=270]{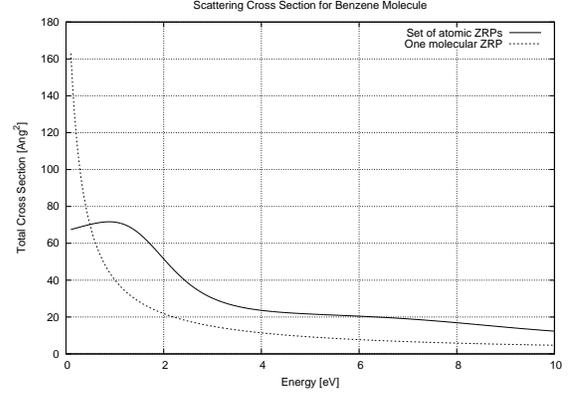}\caption{\label{fig:sigma_R02} Total scattering cross-sections for atomic
and molecular ZRPs: bigger radius $R_{02}$ of benzene molecule (effective
upper estimation).}

\end{figure}

\par\end{center}%
\end{minipage}

Also we notice that better correspondence of the molecular ZRP results
to those obtained with the set of standard ZRPs can be achieved if
in the dressing procedure one includes only the ground state and the
state with maximal energy, i.e. $E_{j}=-\frac{\hbar^{2}\kappa_{j}^{2}}{2\mu}$,
$j=0,\,3$. This can be viewed as if we follow the described general
procedure of application DT for every level, but for each doubly degenerated
level with $E_{j}=-\frac{\hbar^{2}\kappa_{j}^{2}}{2\mu}$, $j=1,\,2$
we perform an additional DT again with the prop function of type \eqref{eq:phi_m}
but choosing a negative transformation parameter $\kappa=-\kappa_{j}$,
$j=1,\,2$ (corresponding to the same energy level). This results
in cancellation of consecutive contributions of the DTs with opposite
to each other spectral parameters, hence the expression \eqref{eq:beta_eff}
gives\begin{equation}
\beta=\left(\kappa_{0}^{-1}+\kappa_{3}^{-1}\right)^{-1},\label{eq:beta_eff_degen}\end{equation}

This result is not surprising since, as it can be seen from the scattering
matrix formalism, such pair of DTs yields consecutive adding and removal
of a pole on the imaginary positive $k$ half-axis (which corresponds
to a bounded state) of \eqref{eq:Darboux_s_l}: $s_{0}\cdot\frac{k+i\kappa_{j}}{k-i\kappa_{j}}\cdot\frac{k-i\kappa_{j}}{k+i\kappa_{j}}=s_{0}$,
$j=1,\,2$. 

That is effective characteristic involves only the lowest and the
highest energy level parameters. In case of odd number of atoms this
would involve only the ground state energy parameter $\kappa_{0}$.

Refined in this way molecular potential gives results demonstrated
on the Fig. \ref{fig:sigma_R01_degen}, \ref{fig:sigma_R02_degen}.

\begin{minipage}[t]{0.45\columnwidth}%
\begin{figure}[H]
\includegraphics[scale=0.3, angle=270]{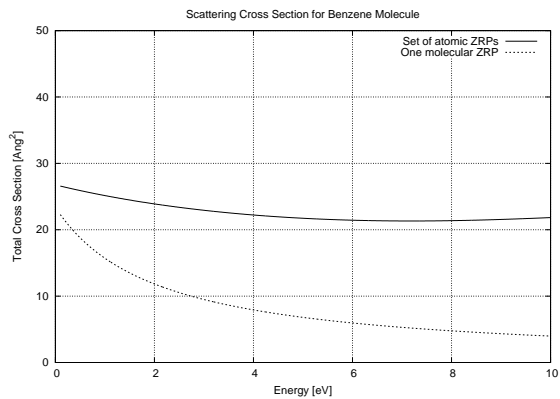}\caption{\label{fig:sigma_R01_degen} Total scattering cross-sections for atomic
and molecular ZRPs: \protect \\
smaller radius $R_{01}$ of benzene molecule, level degeneracy
is taken into account.}

\end{figure}
\end{minipage}$\qquad$%
\begin{minipage}[t]{0.45\columnwidth}%
\begin{figure}[H]
\includegraphics[scale=0.3, angle=270]{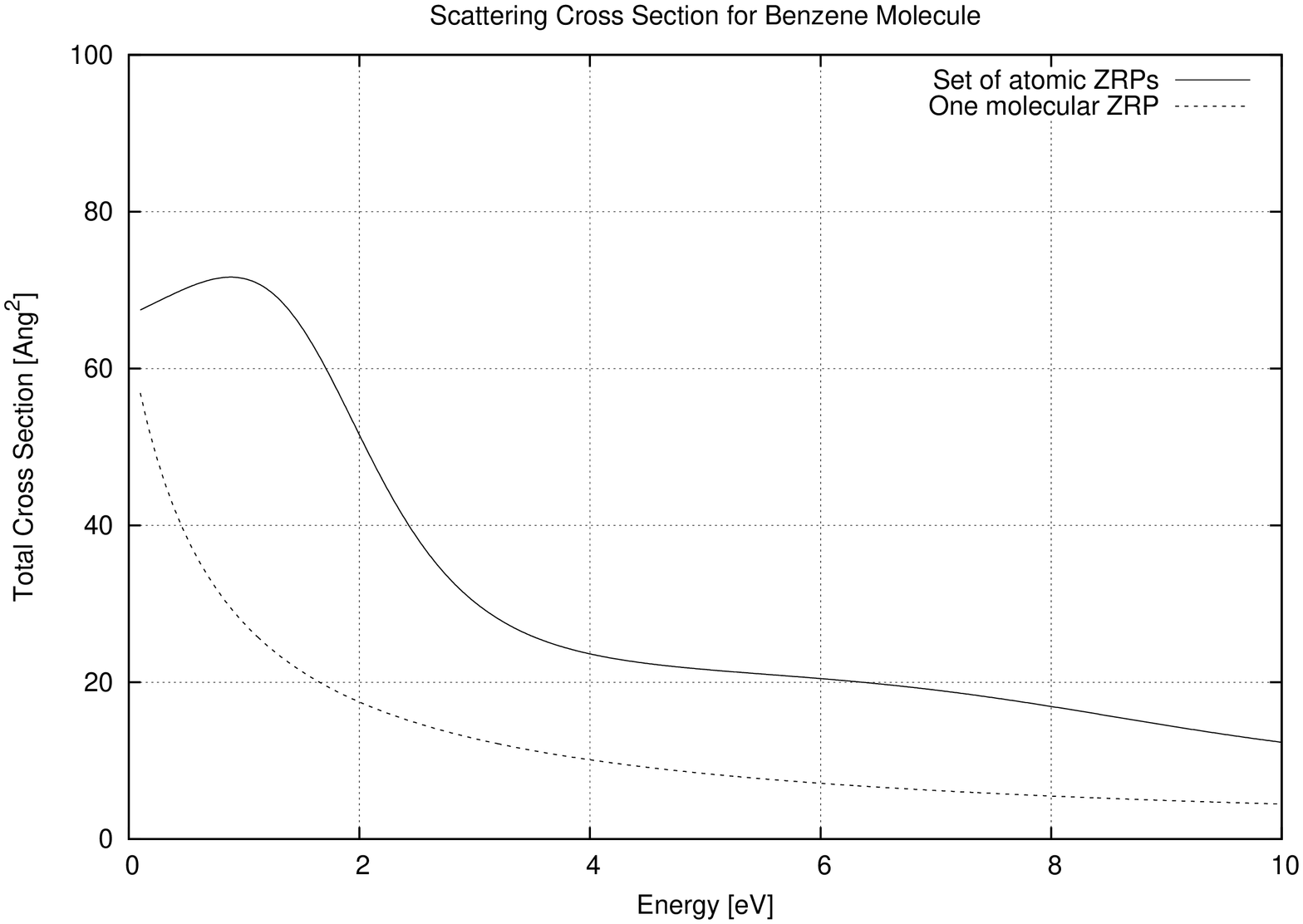}\caption{\label{fig:sigma_R02_degen} Total scattering cross-sections for atomic
and molecular ZRPs:\protect \\
bigger radius $R_{02}$ of benzene molecule, level degeneracy is
taken into account.}

\end{figure}
\end{minipage}

\vspace{15mm}

Below is the table with values of parameters used for calculations
of data presented on the Fig. \ref{fig:ZRP_Huck_comp} - \ref{fig:sigma_R02_degen}.

\vspace{5mm}

\begin{center}
\begin{tabular}{c|c|c|c|c}
$\beta$ & $E_{0}$ & $A$ & $R_{01}$ & $R_{02}$\tabularnewline
\hline
\hline 
$1.5$ $\text{Ang}^{-1}$ & $-7.1$ eV & $0.8$ eV & $1.4$ Ang & $3.6$ Ang\tabularnewline
\end{tabular}
\par\end{center}

\begin{center}
\textbf{\small Table 1.}{\small{} Set of parameters for data plotted
on the figures.}
\par\end{center}{\small \par}

\vspace{5mm}

\section{Conclusions}

The presented method of extension of conventional atomic ZRP approach
is made to describe molecular structures based on their discrete spectrum.
Bounded state energies, either found from conventional ZRP calculations
or in whatever alternative way, determine effective characteristic
of newly introduced single ZRP that is expected to adequately describe
properties of a molecule.

Certainly, replacement of arbitrary molecules with spherically symmetric
point-centers is a crude approximation valid only in low-energy limit,
however, as it was demonstrated, even in case of highly anisotropic
molecules, like benzene, the model gives not perfect but reasonable
quality results at the same time allowing significant simplification
in formulation of more complicated molecular structures problems with
molecules instead of atoms as ZRP constituents. Although we believe
that the molecular ZRP method should be tested on more examples in
order to prove its quantitative merit, it seems to be a beneficial
tool at least to obtain estimates of results on physically reasonable
analytical grounds.

\pagebreak{}

\end{document}